\documentclass[12pt,preprint]{aastex}

\newdimen\minuswidth    %define @ width of minus sign for tables
\setbox0=\hbox{$-$}
\minuswidth=\wd0
\catcode`@=\active
\def@{\kern\minuswidth}
\newdimen\digitwidth    %define ! a one digit width for tables
\setbox0=\hbox{\rm0}
\digitwidth=\wd0
\catcode`!=\active
\def!{\kern\digitwidth}

\newcommand{\mbol}{\ensuremath{M_{\rm bol}}}
\def\reference#1\par{\parindent0pt\hangindent20pt\hangafter1 #1\par}

\received{}
\begin{document}
\shorttitle{Empirical Mass Loss Law}
\shortauthors{Origlia et al.}

\title{DUST IS FORMING ALONG THE RED GIANT BRANCH OF 47 TUC$^1$}
\altaffiltext{1}{
This work is based on observations made with the Spitzer Space Telescope, 
which is operated by the Jet Propulsion Laboratory, 
California Institute of Technology under a contract with NASA.
Support for this work was provided by NASA through 
an award issued by JPL/Caltech.}

\author{Livia Origlia$^2$, Robert T. Rood$^3$, Sara Fabbri$^4$, 
Francesco R. Ferraro$^4$, 
Flavio Fusi Pecci$^2$, R. Michael Rich$^5$, Emanuele Dalessandro$^4$} 
\altaffiltext{2}{INAF--Osservatorio Astronomico di Bologna,
Via Ranzani 1, I--40127 Bologna, Italy, livia.origlia@oabo.inaf.it,
flavio.fusipecci@oabo.inaf.it}
\altaffiltext{3}{Astronomy Department, University of Virginia, Charlottesville, VA 22903,
rtr@virginia.edu}
\altaffiltext{4}
{Universit\`a degli Studi di Bologna, Dip. di Astronomia,
Via Ranzani 1, I--40127 Bologna, Italy,
sara.fabbri@studio.unibo.it,francesco.ferraro3@unibo.it,emanuele.dalessandr2@unibo.it}
\altaffiltext{5}{
Department of Physics and Astronomy, University of California
at Los Angeles, Los Angeles, CA 90095-1547,
rmr@astro.ucla.edu}
%\medskip

\begin{abstract}
We present additional evidence that dust is really forming along the
red giant branch (RGB) of 47~Tuc at luminosities ranging from above
the horizontal branch to the RGB-tip \citep{ori07}. The presence of
dust had been inferred from an infrared excess in the $(K-8)$ color,
with $K$ measured from high spatial resolution ground based near-IR
photometry and ``8'' referring to Spitzer-IRAC 8\micron\ photometry. We
show how $(K-8)$ is a far more sensitive diagnostic for detecting
tiny circumstellar envelopes around warm giants than colors using only
the Spitzer-IRAC bands, for example the $(3.6-8)$ color used by
\citet{boy10}. In addition, we also show high resolution HST-ACS 
$I$ band images of the giant stars which have $(K-8)$ color
excess. These images clearly demonstrate that \citet{boy10} statement
that our detections of color excess associated with stars below the
RGB-tip arise from blends and artefacts is simply not valid.

\end{abstract}


\section{Introduction}
\label{intro}

In \citet{ori07} we presented estimated mass loss rates for first
ascent red giant branch (RGB) stars in the globular cluster 47~Tuc. These
were based on near and mid-infrared photometry obtained from a
Spitzer-IRAC (3.6, 4.5, 5.6 \& 8\,\micron) survey and ground-based
high resolution $JHK$ observations.  We found about 100 giants with ($K-8$) color
excess that we attributed to the presence of dusty circumstellar
envelopes.  These candidate dusty stars were mainly found in the inner
2\arcmin\ (in radius) and had luminosities ranging from above the level of 
the Horizontal Branch (HB) to the Red Giant Branch (RGB)-tip.  For a
given luminosity only a fraction of stars exhibited this color excess,
and this fraction increases toward the tip of the RGB.  From the color
excess we derived a mass loss rate and found a shallower dependence on
luminosity than that expected from Reimers \citep{rei75a,rei75b}
formula. After correcting the observed frequency of dusty envelopes
for incompleteness, we derived an average duty cycle which could be
used along with the observed mass loss rates and evolutionary times to
compute the total amount of mass lost on the RGB.

Recently, \citet{boy10}, using the $(3.6-8)$ Spitzer-IRAC color as
their main diagnostic tool, found evidence for a dust excess only in
asymptotic giant branch stars and possibly a few giants near the RGB tip. On
the basis of this finding, they concluded \citep[see also
][]{boy08,boy09} that our candidate dusty giants below $M_{\rm bol} =
-2.5$ were spurious, mainly blends and/or artefacts.  We emphasize
that {\it i}) this conclusion was not based on a direct star-to-star
check, {\it ii}) their analysis employed a different diagnostic, and
it was not optimized to cover the innermost region of the cluster
sampled by our work.  The AKARI analysis of 47~Tuc presented by
\citet{ita07} also showed dust excess only near the tip, but it had
not the spatial resolution to properly investigate
the stellar population in central regions of globular clusters.  We
ourselves using ISOCAM found circumstellar dust excess only near the
RGB tip \citep{ori02}, again because of the lower spatial resolution
of ISOCAM compared to IRAC.
 
In this Letter we respond to the \citet{boy10} criticism. We demonstrate
the importance of an optimum sampling of the innermost region of the
cluster where most of the red giants, along with the rest of the
stellar population, are found. We show that $(K-8)$ is a much better
diagnostic of dusty giants than $(3.6-8)$. Finally, using a high
resolution HST-ACS $I$ band image, we clearly demonstrate that our
detected $(K-8)$ color excess in giant stars below the RGB tip is not
an effect of blending and/or an artefact.

\section{Photometric Samples}

Many of the contentions of \citet{boy10} arise from their
misunderstanding of the ``shallow'' and ``deep'' samples
of \citet{ori07}. It is clearly stated in Secton 2 and in the caption of
Figure~2 of \citet{ori07} that shallow and deep do not refer to the
short and long Spitzer exposures.  Both the shallow and deep samples
include photometry from short (0.6 s) and long (12 s) exposures.  We
use the very shortest exposures to measure the brightest stars near
the RGB-tip, which are saturated in the long exposures. The long
exposures are averaged to study the fainter stars. The shallow
sample refers to a $5\arcmin \times 8\arcmin$ rectangular area
photometrically cut at $K=11$ (or $M_{\rm bol} \sim 0$), since at
fainter magnitudes crowding severely affects the photometric accuracy
in the central region.  The deep sample down to $K\approx 14$ excludes the central 2\arcmin\
in radius.

The Rood-shallow and Rood-deep samples of \citet{boy10} have no
correspondence with the shallow and deep samples of \citet{ori07},
either in terms of the sampled region or in terms of total exposure
time.  The \citet{boy10} Rood-deep sample includes the very center of
the cluster, so they find more scatter. They do discuss completeness
of their samples, but in detecting color excesses scatter is more
important than completeness. We feel that their sampling of the
central part of the cluster is adequate to detect only the brightest
dusty stars. They do discuss the region with $R>3\arcmin$ separately
but that does not correspond to our deep sample at $R>2\arcmin$.  Their deep
samples in the $3\arcmin <R<10\arcmin$ range have small overlap with the
$5\arcmin \times 8\arcmin$ field of view sampled in \citet{ori07}.  We
have examined our sub-sample at $R>3\arcmin$ and find that only one dusty
AGB star has been detected, fully consistent with the \citet{boy10}
finding in this outer region.

\section{Diagnostic Planes}

\citet{boy10} claim that the  $M_{\rm bol},~(3.6-8)$
diagnostic plane that they use is as effective in detecting dusty
giants as the $M_{\rm bol},~(K-8)$ plane used by \citet{ori07}. 
Other than stating that dust emission at 3.6\micron\ is negligible,
they do not provide any direct test of their supposition.

Figures~1 \& 2 show $\mbol,~(K-8)$ and $\mbol,~(3.6-8)$ color-magnitude
diagrams of the shallow and deep samples from \citet{ori07}
photometry.  Triangles mark those stars with a 3$\sigma $ excess
in the $(K-8)$ color. Circled triangles show those which also have a
3$\sigma $ excess in the $(3.6-8)$ color.

Figure 3 shows the $(K-8)$,~$(3.6-8)$ color-color diagram.
Triangles marks those stars showing a 3$\sigma $
excess in the $(K-8)$ color. Filled triangles show those which also have a 
3$\sigma $ excess in the $(3.6-8)$ color. 

The figures all show that the $(K-8)$ color is far more effective than
the $(3.6-8)$ color in disentangling relatively warm ($T_{\rm
eff}>4000$ K) photospheres from optically thin, warm ($T_{\rm
dust}>500$ K) dusty envelopes.  Most of the candidate dusty stars
identified by \citet{ori07} have $(3.6-8)$ colors in the 0.1--0.3 mag
range, barely (if at all) exceeding the 3$\sigma$ photometric
uncertainty which ranges from 0.15 at $M_{\rm bol}<-2$ to 0.25 at
$M_{\rm bol}\approx 0$.  The $(K-8)$ color always exceeds 0.2 mag and
spans a larger range.  
We recall that, according to the simulations with the
DUSTY code, these $K-8$ and $3.6-8$ colors are those of a
dust that is optically thin ($\tau_8$ is in the 0$^{-4}$-10$^{-2}$ range)
and made of silicates, i.e., a chemical composition typical of O-rich
environments as the circumstellar envelopes of low mass giants.  Such a dust has
IRAC-Spitzer color temperatures in the 400-800 K range.

In relatively warm and low luminosity giants,
like low mass RGB stars, the fractional contribution of the warm dust
emission to the 3.6\micron\ is {\em not} negligible and {\em not} much
smaller than the dust contribution at 8\micron.  Indeed, in the
measured giants of 47~Tuc we estimate that the average dust
contributions are ~20\% at 3.6\micron\ and ~30\% at 8\micron.  The
fractional contribution of dust emission at 3.6\micron\ is negligible
only in much cooler and more luminous giants stars and/or in case of
envelopes with a large amount of dust.

Hence, when using the $(3.6-8)$ color to select candidate dusty RGB
stars in globular clusters, as done by \citet{boy10}, one is clearly
biased towards the coolest (hence the most luminous) ones and with
relatively large amount of circumstellar dust.

\section{Blends and Artefacts}

Since the IRAC-Spitzer pixel is relatively large, it is possible that
more than one star actually falls in it.  Hence blending is an obvious
worry near cluster centers, although only a rather special kind can
mimic a dusty star. If more than one star blends together in the
8\micron\ PSF, and the blending object does not fall in the $K$ or
3.6\micron\ PSF a bogus IR excess can result.  
However, we emphasize that to produce an
appreciable excess the blending star(s) must be comparable in
brightness (say within a factor of 10) to the first star.
More precisely, the 3$\sigma$ color excess on the upper RGB is
$(K-8)>0.15$. The blending stars would require 8\micron\ luminoisities within a
factor of roughly 6 (2 mag) to produce a significant excess. On the
lower RGB the 3$\sigma$ color excess is $(K-8)>0.15$, and the luminosities
would have to be within a factor of 4 (1.3 mag).
Given that the RGB of a globular cluster is practically vertical in
the $K,(K-8)$ color-magnitude diagram (CMD) (i.e., the same $K-8$ 
color for normal giants),
differences in luminosity within factors of 4--6 are also
required in the K band.  The corresponding I-band luminosites should
vary only by factors of 2.5--4 (i.e stars cannot differ by more than
1--1.5 mag in the lower and upper RGB, respectively).

In \citet{boy10} and earlier papers \citep{boy08,boy09} on different
clusters, the authors have made much the possibility of blending and
argued that the bulk of the the dusty stars in \citet{ori07} are
blends.  We stress that their statement is not based on a direct
inspection of the stars in our sample, but simply on the appearance of
their CMDs in a different photometric plane, in different sampled
regions or even in different clusters.  To estimate the possible
magnitude of the blending problem, \citet{boy10} employ a
de-convolution technique in which the 3.6\micron\ images are analyzed
using the 8\micron\ PSF.  While this technique might give some estimate
of the magnitude of blending problems it is definitely less effective
than a direct inspection of high spatial resolution optical/near IR
images as done in \citet{ori07}.

In order to definitely solve the issue of blending we have obtained an
HST ACS $I$ band image of 47~Tuc from the archive. Figure~\ref{ima4}
shows 78 of of 93 candidate dusty giants.  For the majority (72 out of
78) of stars (top panel in Figure \ref{ima4}), only the target and a
few (if any) significantly fainter stars are present within the
IRAC@Spitzer PSF area. These stars, which are typically at least 2--3
magnitudes fainter, cannot be responsible for the observed color
excess.  The only possible exceptions are ID-351, 373 and 489, where
the neighboring stars are $\sim 1$ magnitude fainter.  ID-351 is also
a blend in the near IR $K$ images, so at least one of the two
components (probably the brighter one) is likely responsible for the
observed color excess.  ID-373 and 489 both have a color excess
exceeding $\sim0.6$ magnitude, which is too large to be accounted for by a blending
star $\sim 1$ magnitude fainter.

Among the remaining 6 candidate stars (bottom panel in
Figure \ref{ima4}) with color excess, ID-76 and ID-103 are only 
marginally blended, while the 2 stars blending
ID-240 are not bright enough to produce the detected excess.  ID-445,
454 and 479 are a blend of 2 stars with similar luminosity both in the
near IR an Spitzer PSFs. In this case we cannot precisely identify the
star(s) responsible for the color excess but at least one has it.  

The other 15 candidate dusty giants not imaged by ACS lie in the outer
region of the field of view covered by our Spitzer and near IR surveys, where
crowding is less of a problem.  Indeed, 12 are definitely free from
blends by stars bright enough to produce a detectable excess, while 3
(2 of which are AGB stars) are only partially blended. 

Hence, just as we described in Section 2 of \citet{ori07}, a suitable cross
correlation with high spatial resolution optical and/or near infrared stellar
catalogs allows for 1) a proper identification of the stellar
counterpart, 2) the correct determination of the star photospheric
parameters, and 3) a direct way to check for and remove possible
blends, artefacts, background galaxies etc.  It is also worth noting
that \citet{ori07} used ROMAFOT \citep{buo83}, a software package
optimized to perform PSF fitting photometry of crowded fields even in
a regime of under-sampling, as with Spitzer-IRAC at the shortest
wavelengths.  ROMAFOT allows one to directly inspect the fit of faint
and/or problematic sources to double check possible residual blending,
artefacts, background galaxies etc. and to remove them.  Objects
identified as blends have been rejected as dusty giants.  
Blending turned out to be a rare event (P$<$ a few percent in the lower
RGB and P$\simeq$0 in the upper RGB) in mimicking color excess, as
expected.

Another indication that blending is not a problem is the
absence of any correlation between the occurrence of blending
and luminosity in the sampled RGB portion.
Figures 1 \& 2 also show that most of the candidate dusty giants are
confined within the central 2\arcmin. This might be taken as an
indication of blending, but it turns out that it is just what is
expected. If the dusty stars are real they should be distributed like
the cluster light. False dusty stars arising from crowding should be
preferentially found in regions of high density. Using a King  model 
of the cluster, we find that of the total bolometric 
light sampled in the central 
2\arcmin\ (in radius) about 64\% is contained in the central 1\arcmin\ 
and 36\% in the annulus between 1\arcmin\ and 2\arcmin. 
The corresponding fractional number of dusty giants in the same regions are 
67$\pm$11\% and 33$\pm$7\%. 
Very similar results within the errors (i.e. 61\%$\pm$5\% and 39\%$\pm$4\%, respectively) 
are obtained by counting non-dusty stars.
The observed numbers are completely consistent with model expectations and this 
should be the case if crowding is not a problem. 

\citet{boy10}, relying on Spitzer data alone, may be bothered by
instrumental artefacts.  We have no such problem since our high
spatial resolution optical and near IR data have no (or at least
different) artefacts.

In summary, \citet{boy10} conclusion that \citet{ori07} candidate dusty RGB
stars below the tip are false detections is not justified.

\section{Discussion and Conclusions}

The $(K-8)$ color excess detected by \citet{ori07} in a fraction of RGB
stars down to $\mbol \sim 0$ within the central 2\arcmin\ 
of 47~Tuc is real. It is not an artefact, since blends and other
possible spurious effects have been properly accounted for by
cross-correlating the Spitzer sources with high spatial resolution
optical and near infrared catalogs of stellar counterparts.

The fact that \citet{boy10} did not find as many dusty stars in the
central 2\arcmin\ is mainly due to the use of a different diagnostic,
namely the $(3.6-8)$ color, which is only effective for detecting dust
excess in cooler (hence more luminous) giants. Further, their
photometric analysis is not optimized in the central, densest region.
\citet{boy10} find only few dusty giants in the outer regions, fully
consistent with \citet{ori07} and the additional CMDs shown in Figure
2.

As discussed in Section 4 of \citet{ori07}, mass loss rates and duty
cycles decrease with decreasing stellar luminosity.  For a given
luminosity, the estimated rates are higher (between a factor of 2 near
the tip up to 2 orders of magnitude down to $\mbol \sim 0$ than those
predicted by the Reimers law, as also noted by \citet{boy10}. The
shallower slope we found has been also suggested by \citet{mes09},
from chromospheric line wind diagnostics of giants in metal poor
clusters.  The Reimers result was obtained from Population~I objects,
and one might anticipate some differences.  Our higher mass loss rate
is not a major problem, given that both the Reimers law and our
results contain free parameters, {\it i}i) the Reimers efficiency
$\eta$, and {\it ii}) the gas to dust ratio and expansion velocity. 
In either case the
free parameters must be set by indirect observational constraints on
the total mass lost during the RGB evolution like the HB
morphology. 

The total mass loss occurs predominantly in the upper $\sim2$
magnitudes near the RGB tip. The contribution of the low luminosity
giants ($\mbol > -1$) is small ($<20$\%) to negligible, and within the
estimated uncertainty.  The importance of our observation of mass loss
in less luminous stars is not its impact on total mass lost, but
rather the clue it gives us to the physics of mass loss. Likewise, the
episodic nature of mass loss in non-variable stars tells us
something. In both cases, it seems quite clear that the underlying
driver of the mass loss is {\em not} radiation pressure on
dust. Another important outcome of our larger project will be an
investigation of the differential mass loss among clusters with
different metallicities and HB morphologies.  This has been the goal
of our Spitzer survey and the complementary ground based observations,
and the results will appear in forthcoming papers.

\acknowledgements This research was supported by the Agenzia Spaziale  
Italiana
(under contract ASI-INAF I/016/07/0)  and by the Ministero  
dell'Istruzione,
dell'Universit\`{a} e della Ricerca.

\begin{figure}
%\epsscale{1.8}
\plotone{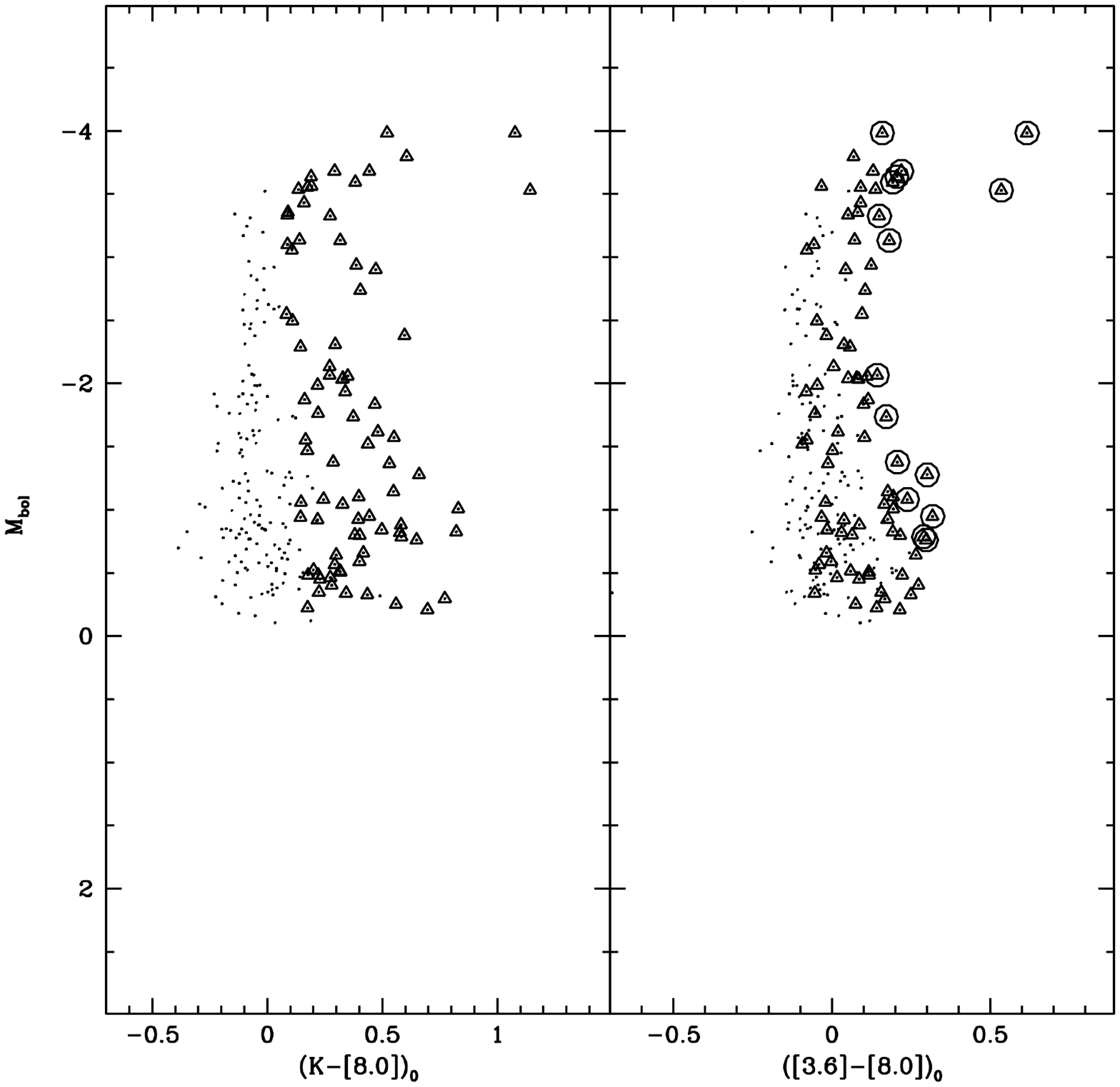}
\caption{$\mbol,~(K-8)$ (left) , $\mbol,~(3.6-8)$ (right) CMDs of the
47 Tuc shallow sample of \citet{ori07}, i.e. within a rectangular
region of $5\arcmin \times 8\arcmin$ around the cluster center.
Triangles are stars with a 3$\sigma$ ($(K-8)$) color excess. Circled
triangles are stars which also have a 3$\sigma$ $(3.6-8)$ color excess.
\label{ima1}}
\end{figure}

\begin{figure}
%\epsscale{1.8}
\plotone{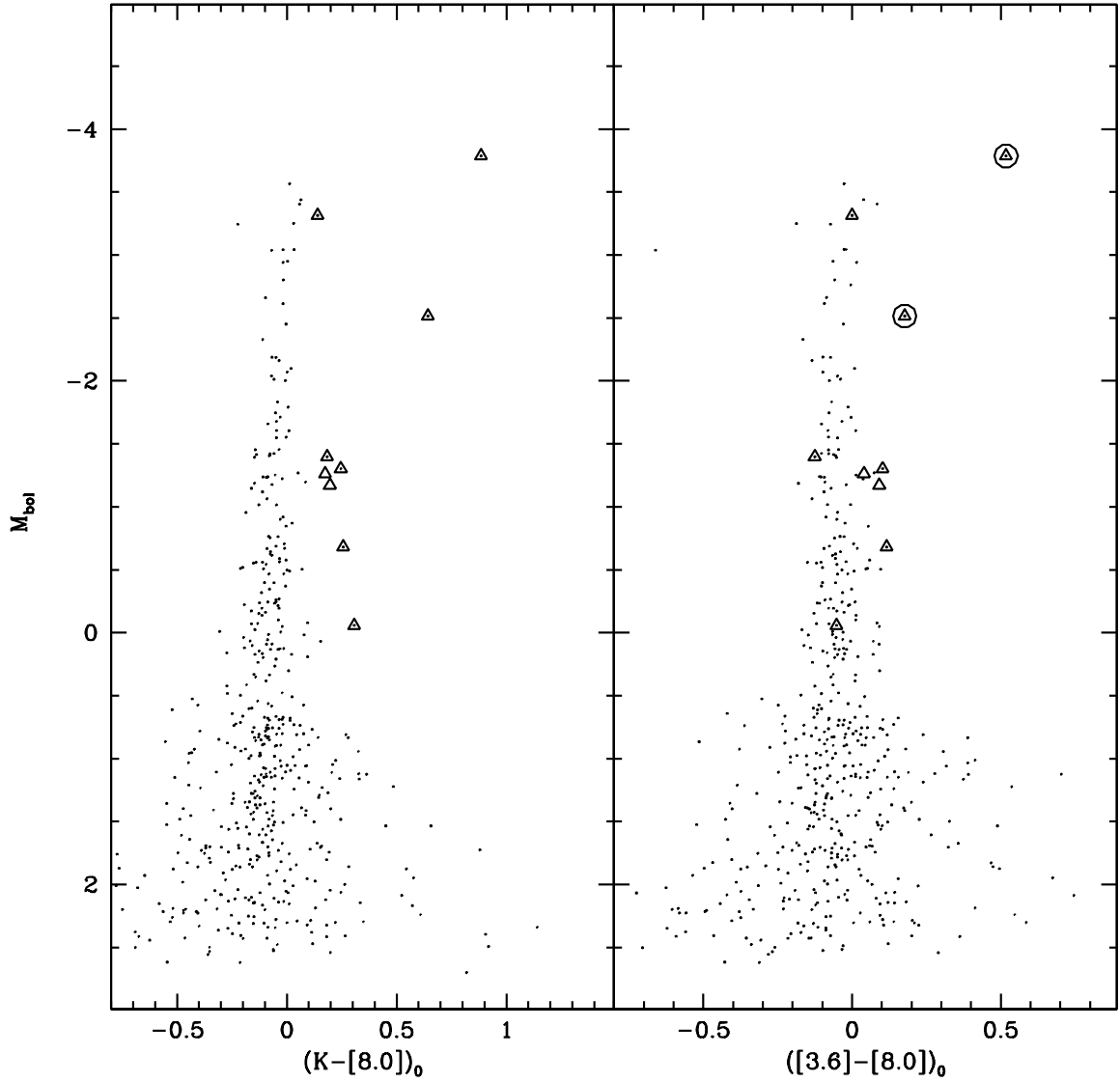}
\caption{$\mbol,~(K-8)$ (left) , $\mbol,~(3.6-8)$ (right) CMDs of the
47Tuc deep sample of \citet{ori07}, i.e. within a rectangular region
of $5\arcmin \times 8\arcmin$ around the cluster center, but excluding
the central region, $r<2\arcmin$. Triangles are stars with a
3$\sigma$ $(K-8)$) color excess. Circled triangles are stars which
also have a 3$\sigma$ $(3.6-8)$ color excess.
\label{ima2}}
\end{figure}

\begin{figure}
%\epsscale{1.8}
\plotone{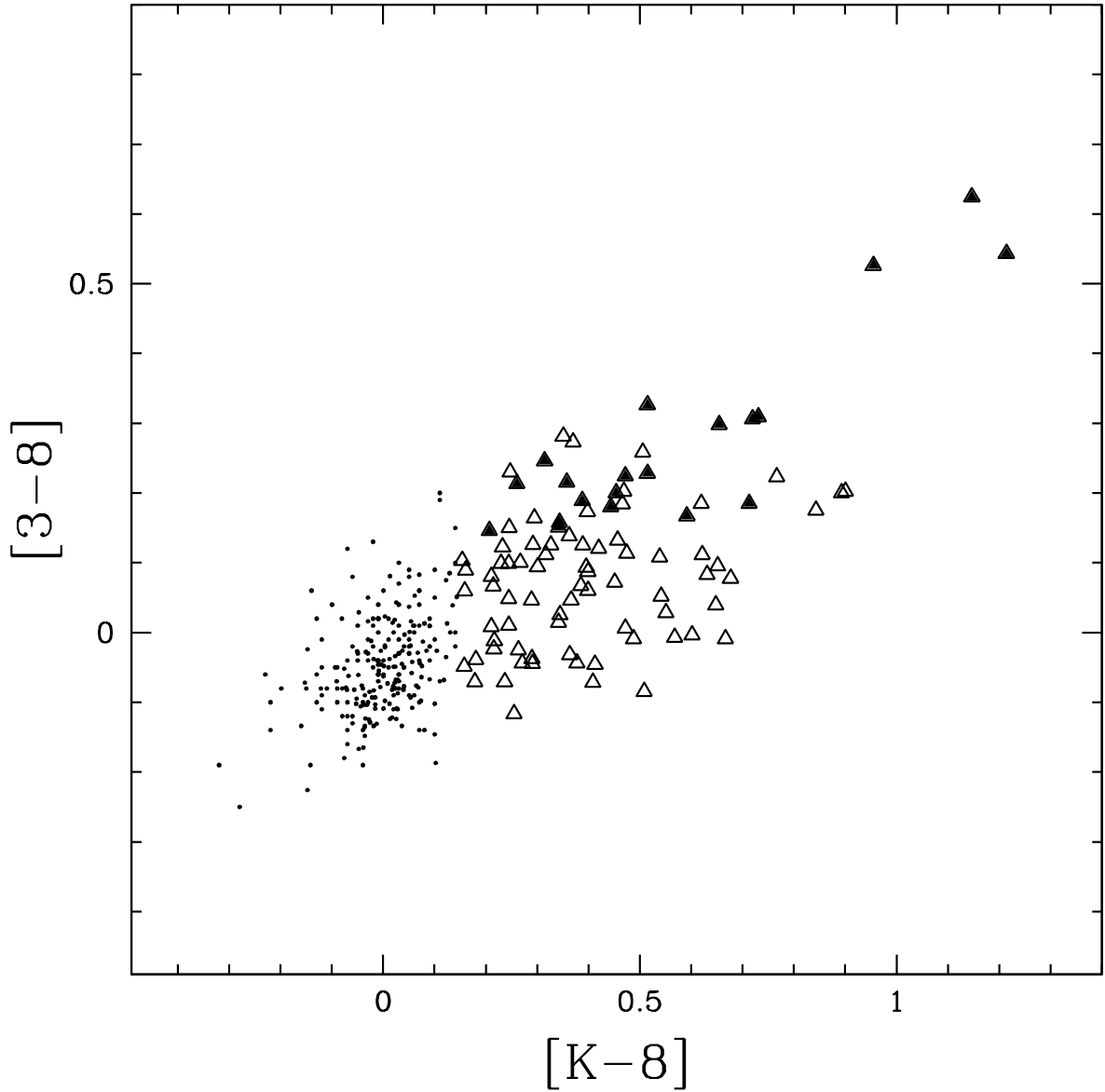}
\caption{($K-8$),~($3.6-8$) color-color diagram of 47Tuc within a
rectangular region of $5\arcmin \times 8\arcmin$ around the cluster
center.  Triangles are stars with a 3$\sigma$ ($(K-8)$) color
excess. Filled triangles are stars which also have a 3$\sigma$ $(3.6-8)$
color excess.
\label{ima3}}
\end{figure}

\begin{figure}
\epsscale{0.9}
\plotone{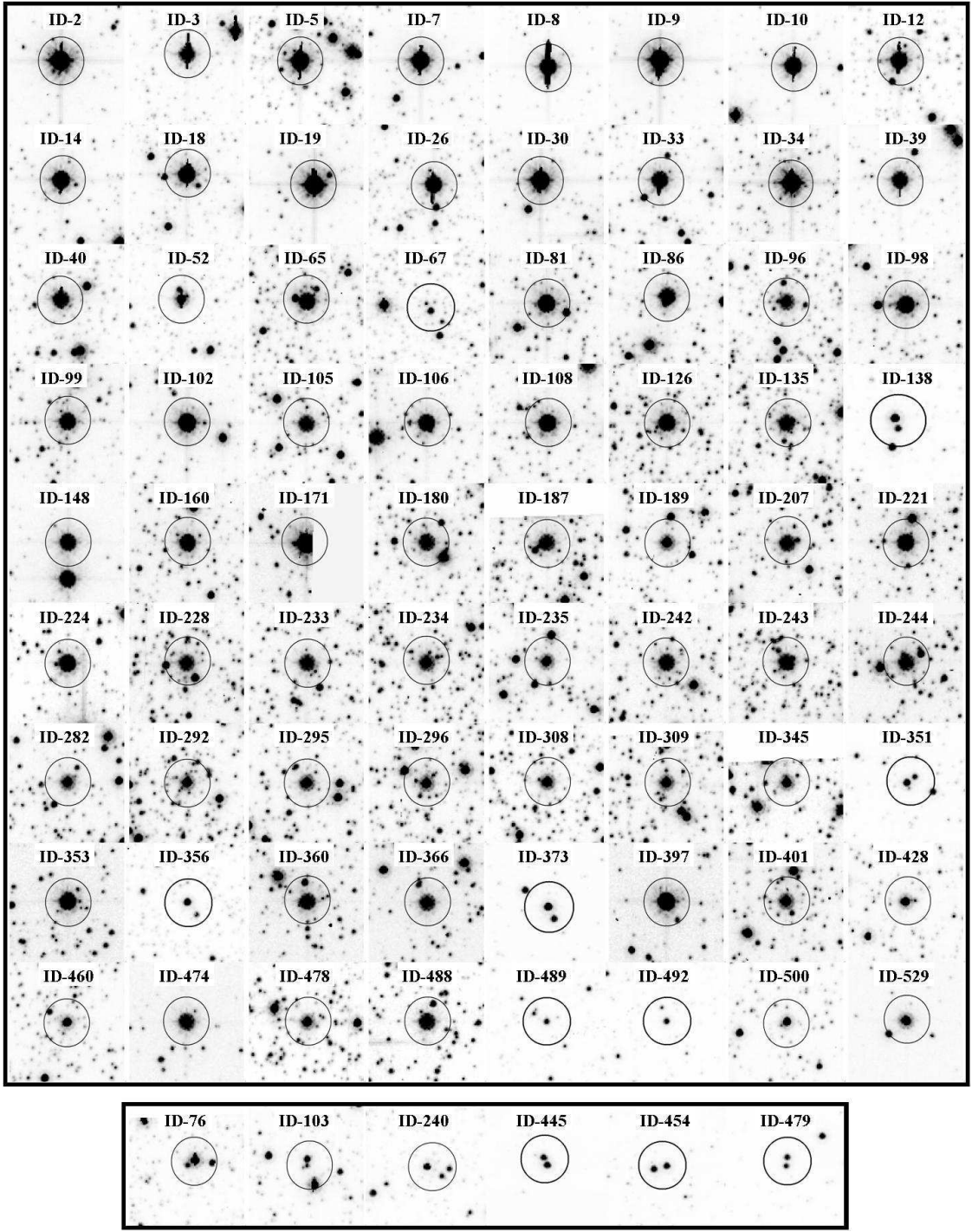}
\caption{ACS $I$ band images of 78 candidate dusty stars in 47
Tuc. The field of view of each image is $5\arcsec\times 5\arcsec$ and
the circles ($\approx1$\arcsec in radius) mark the 8\micron\ PSF area
of IRAC.
\label{ima4}}
\end{figure}

\end{document}